\documentclass[%
reprint,
superscriptaddress,
amsmath,amssymb,
prb,
]{revtex4-1}

\usepackage{amsmath} % AMS Math Package
\usepackage{amsthm} % Theorem Formatting
\usepackage{amssymb}	% Math symbols such as \mathbb
\newcommand{\ket}[1]{\left| #1 \right>} % for Dirac bras
\newcommand{\matrixel}[3]{\big< #1 \big| #2 \big| #3\big>} % for Dirac kets
\newcommand{\spinT}[2]{\left| #1 \right> \leftrightarrow \left| #2\right>} % for Dirac kets
\newcommand{\natSi}{$^\text{nat}$Si} % for natural silicon
\newcommand{\pureSi}{$^{28}$Si} % for natural silicon
\newcommand{\corrparam}{\epsilon_{\text{corr}}}
\newcommand{\correnergy}{E_{\text{corr}}}

\let\baraccent=\= % rename builtin command \= to \baraccent
\renewcommand{\=}[1]{\stackrel{#1}{=}} % for putting numbers above =

\theoremstyle{definition}

\theoremstyle{remark}

\newcommand{\degreeC}{^{\circ}\mathrm{C}} % for degree celsius
\usepackage{natbib}

\usepackage{graphicx}% Include figure files
\usepackage{dcolumn}% Align table columns on decimal point
\usepackage{bm}% bold math
\usepackage{color}

\begin{document}

\preprint{Preprint version: \today}

\title{Hyperfine Clock Transitions of Bismuth Donors in Silicon Detected by Spin Dependent Recombination}% Force line breaks with \\
%\thanks{A footnote to the article title}%

\author{P. A. Mortemousque}
\affiliation{School of Fundamental Science and Technology, Keio University, 3-14-1 Hiyoshi, Kohoku-ku, Yokohama 223-8522, Japan}

\author{S. Berger}
\affiliation{School of Fundamental Science and Technology, Keio University, 3-14-1 Hiyoshi, Kohoku-ku, Yokohama 223-8522, Japan}
\affiliation{Department of Physics, ETH Zurich, CH-8093 Zurich, Switzerland}

\author{T. Sekiguchi}
\affiliation{School of Fundamental Science and Technology, Keio University, 3-14-1 Hiyoshi, Kohoku-ku, Yokohama 223-8522, Japan}

\author{C. Culan}
\affiliation{School of Fundamental Science and Technology, Keio University, 3-14-1 Hiyoshi, Kohoku-ku, Yokohama 223-8522, Japan}

%\author{F. Hoehne}
%\affiliation{Walter Schottky Institut, Technische Universit\"at M\"unchen, Am Coulombwall 4, 85748 Garching, Germany}

%\author{M. S. Brandt}
%\affiliation{Walter Schottky Institut, Technische Universit\"at M\"unchen, Am Coulombwall 4, 85748 Garching, Germany}

\author{R. G. Elliman}
\affiliation{Australian National University, Research School of Physics and Engineering, Canberra, ACT 0200, Australia}

\author{K. M. Itoh}
\affiliation{School of Fundamental Science and Technology, Keio University, 3-14-1 Hiyoshi, Kohoku-ku, Yokohama 223-8522, Japan}

\begin{abstract}
Bismuth donors ion-implanted in $^{28}$Si and $^\text{nat}$Si are studied using magnetic resonance spectroscopy based on spin dependent recombination. 
The hyperfine clock transition, at which the linewidth is significantly narrowed, is observed for the bismuth donors.
% is robust against electric field perturbations 
The experimental results are modeled quantitatively by molecular orbital theory for a coupled pair consisting of a bismuth donor and a spin dependent recombination readout center, including the effect of hyperfine and Zeeman interactions.
\end{abstract}

%\pacs{Valid PACS appear here}% PACS, the Physics and Astronomy
                             % Classification Scheme.
%\keywords{Suggested keywords}%Use showkeys class option if keyword
                              %display desired
\maketitle

%%%%%%%%%%%%%%%%%%%%%%%%%%%%%%%%%%%%
%%%%%%%%%%%%%%%%%%%%%%%%%%%%%%%%%%%%
%%%%%%%%%%%%%%%%%%%%%%%%%%%%%%%%%%%%
\section{Introduction}

Among a variety of physical systems investigated for quantum information processing, superconducting qubits are one of the promising candidates as quantum processors because of their fast operation capabilities and their potential for scalability.\cite{Devoret2013} However, because of their relatively fast decoherence rate which might be insufficient for maintaining quantum information throughout the course of computation, development of quantum memories that could support the operation of the superconducting processors are desired.  Such memory qubits have to be addressable at low magnetic field ($<10$ mT for aluminum\cite{Cochran1958}), since superconducting qubits become unoperable at magnetic fields higher than their critical fields.

Within this context, a bismuth (Bi) donor in silicon (Si) has attracted much attention recently. Its large hyperfine interaction $A=1.4754$ GHz (Ref. \onlinecite{Feher1959}) and the $^{209}$Bi nuclear spin $I=9/2$ give rise to a large zero-field splitting of 7.4 GHz that is comparable to the typical energy splitting between $\ket{R}$ and $\ket{L}$ states of superconducting flux qubits.\cite{Chiorescu2003} 
Thus, coherent coupling between a Bi spin qubit in Si and a superconducting flux qubit on Si is in principle possible via a microwave photon traveling through a waveguide placed between the two qubits. \cite{Morley2010,George2010} 
%In parallel to this, motivated by theoretical proposals,\cite{Marcos2010,Diniz2011} the coupling of electron spins of NV$^-$ centers in diamond with a coplanar waveguide resonator\cite{Kubo2010} and with a flux qubit\cite{Zhu2011} has already been demonstrated.
The proposal to couple Bi in Si with a superconducting qubit\cite{George2010} have triggered extensive fundamental studies of the Bi donor in Si very recently. 
Starting from the spectroscopic analysis of the electron paramagnetic resonance (EPR),\cite{George2010,Weis2012} the electron spin relaxation time $T_1$,\cite{Morley2010,Belli2011} decoherence time $T_2$,\cite{Morley2010,Belli2011,Weis2012,Wolfowicz2012,Wolfowicz2013} and superhyperfine interaction with nearby $^{29}$Si nuclear spins \cite{Belli2011,Balian2012} were investigated. Moreover, the 
coherent coupling between %coherence transfer between 
the Bi electrons and $^{209}$Bi nuclear spins\cite{George2010} and dynamic nuclear polarization of $^{209}$Bi were achieved.\cite{Morley2010,Sekiguchi2010} 
Hybrid nuclear-electronic qubits consisting of superpositions of electronic and nuclear spin states have been used to demonstrate five orders of magnitude longer coherence times than the manipulation times.\cite{Morley2012} 
In order to extend the coherence time of Bi donor electrons, magnetic field-insensitive clock transitions can be used.\cite{Wolfowicz2013,Mohammady2010,Balian2012} Also, at low temperatures, the presence of $4.7\%$ $^{29}$Si ($I$=1/2) in naturally available silicon  ($^{\text{nat}}$Si) limits the coherence time of donors\cite{Abe2010,Witzel2010}  so that the use of isotopically purified $^{28}$Si is helpful.\cite{Tyryshkin2011,Steger2012,Wolfowicz2013}
The fact that most of aforementioned Si:Bi studies were performed in the past three years shows how rapidly developing this field is.  However, one aspect that has been scarcely studied is the investigation of Si:Bi at low-fields to enable the coupling to superconducting qubits.  In order to fill in this gap, we have shown recently \cite{Mortemousque2012} that magnetic resonance spectroscopy with detection based on spin dependent recombination\cite{Lepine1972} (SDR) allows to manipulate and detect spins at  low magnetic fields.

In the present study, using such a capable SDR technique, we perform spectroscopy of bismuth implanted in both $^\text{nat}$Si and isotopically enriched $^{28}$Si samples and observe a significant line narrowing at the hyperfine clock transition (HCT), where the transition frequency $\nu$ is insensitive to the change in $A$ induced by variations in charge distribution ($\partial \nu / \partial A =0$). 
While existence of optimal working points (e. g., gate voltages) at which superconducting qubits are immune to the electric charge noise has been demonstrated,\cite{Vion2002} observation of HCT in solid state systems has never been reported to our knowledge. The HCT is different from the conventional clock transition, which is insensitive to magnetic noise ($\partial \nu / \partial B_z =0$). The conventional clock transitions are routinely employed in the operation of atomic clocks\cite{Lyons1952, Kusch1949} utilizing $^{133}$Cs and trapped ions.\cite{Diddams2001} A similar clock transition of bismuth donors in silicon has been adopted to achieve extremely long donor electron spin coherence time.\cite{Wolfowicz2013} HCT investigated in this study is more involved in the sense that the hyperfine interaction of a donor can be affected by both strain and electric field fluctuations.
Away from the HCT point, the interaction of a donor (D) electron with a nearby implantation defect, which is used in SDR spectroscopy as a readout center (R), causes an asymmetric broadening of the spectral line shapes. 
This interaction is equivalent to an effective electric perturbation. 
Thus we propose a theoretical model that describes the change of the donor wave function due to the presence of this readout center. This model makes it possible to simulate the SDR spectra and estimate the associated change in the hyperfine interaction. Finally, we compare the line position and the line shape measured by SDR spectroscopy with our calculation and extend the theoretical model for other donors in silicon.

%%%%%%%%%%%%%%%%%%%%%%%%%%%%%%%%%%%%
%%%%%%%%%%%%%%%%%%%%%%%%%%%%%%%%%%%%
%%%%%%%%%%%%%%%%%%%%%%%%%%%%%%%%%%%%
\section{Experimental observation of the change in hyperfine interaction}

%%%%%%%%%%%%%%%%%%%%%%%%%%%%%%%%%%%%
%%%%%%%%%%%%%%%%%%%%%%%%%%%%%%%%%%%%
	\subsection{Samples}
Two types of samples were employed; a silicon crystal enriched to 99.983\% $^{28}$Si ($[^{29}\text{Si}]=90$ ppm and $[^{30}\text{Si}]=80$ ppm) with a resistivity $\approx10$ $\Omega \cdot$cm and a highly resistive ($>3$ k$\Omega \cdot$cm) float-zone $^{\text{nat}}$Si. 
These two substrates were ion-implanted with Bi and are labeled \pureSi :Bi and \natSi :Bi, respectively. The ion implantations were performed at room temperature with the total fluence of $2\times10^{13}$ cm$^{-2}$. The implantation energies were 300 and 550 keV with the doses of $0.7\times10^{13}$ and $1.3\times10^{13}$ cm$^{-2}$, respectively.  These conditions yielded a maximum bismuth concentration of $1.8\times10^{18}$ cm$^{-3}$ (above the solubility limit\cite{Trumbore1960}) in the depth of 90 to 150 nm from the surface. The post-implantation annealing, performed at 650 $\degreeC$ for 30 min in an evacuated quartz tube, led to an activation efficiency\cite{Marsh1968,Baron1969,deSouza1993,Weis2012}  below $60\%$, resulting in the Bi donor concentration less than $1.1\times10^{18}$ cm$^{-3}$ (below the metal-insulator transition\cite{Abramof97}). This process was designed to maximize the number of D-R pairs, instead of fully activating all the implanted Bi atoms.\cite{Studer2013}

%%%%%%%%%%%%%%%%%%%%%%%%%%%%%%%%%%%%
%%%%%%%%%%%%%%%%%%%%%%%%%%%%%%%%%%%%
	\subsection{SDR method}

The continuous illumination provided by a 100-W halogen lamp (above band-gap power of 100 mW/cm$^2$ outside the EPR cavity) generated photoexcited electrons in the sample. 
The capture of photocarriers by the ionized donors of D-R pairs takes place on a time scale $\tau_{ec}$ of the order of 10 to 100 $\mu$s for an illumination at 635 nm of 20 mW/cm$^2$ at 5 K.\cite{Hoehne2013-2} 
For the phosphorus donor coupled to a dangling bond readout center, the expected recombination time for the antiparallel electron spin pair was typically $\tau_\text{ap}\approx10$ $\mu$s whereas for the parallel spin pair, the recombination time $\tau_\text{p} \approx 1$ ms was much longer.\cite{Hoehne2013-2}  
%Also, preliminary time-resolved electrically detected magnetic resonance (EDMR) measurements of Bi-R pairs in \natSi :Bi showed a dynamics similar to the phosphorus donor coupled to a dangling bond.
Preliminary time-resolved electrically detected magnetic resonance (EDMR) measurements of Bi-R pairs in \natSi :Bi showed a dynamics similar to the donor coupled to a dangling bond defect situating at the Si/SiO$_2$ interface even though the readout centers R created by the implantation were situated around 90 nm deep Bi donors.
%Preliminary time-resolved electrically detected magnetic resonance (EDMR) measurements of Bi-R pairs in \natSi :Bi showed a dynamics similar to the phosphorus donor coupled to a dangling bond even though these dangling bonds are situated at the Si/SiO$_2$ interface whereas the readout center R are created by the Bi-ion implantation. 
%Even though the nature of the donors and of the dangling bonds at Si/SiO2 interface used in Ref. \onlinecite{Hoehne2013-2} and the readout center R created by Bi-ion implantation are different, preliminary time-resolved electrically detected magnetic resonance (EDMR) measurements of Bi-R pairs in natSi:Bi showed a dynamics similar to the phosphorus donor coupled to a dangling bond.
%Since the recombination time for the parallel ($\tau_p$) case was much longer than that for the antiparallel ($\tau_{ap}$) case, only the parallel spin pairs remained in the steady state illumination without external induction of the magnetic resonance. 
As a consequence, only the parallel spin pairs remained in the steady state under illumination without external induction of the magnetic resonance. 
Therefore, flipping the donor electron spins by the external magnetic resonance irradiation broke this steady-state constant current situation and decreased the photocurrent by the enhancement of the spin-dependent recombinations.\cite{Lepine1972} Such a  change of the sample photoconductivity led to a decrease in the absorption of the microwave electric field by the sample (photocarriers) leading to an enhancement in the $Q$ factor of the EPR cavity. 
The defect utilized as a readout center in this study had a $g$ factor of $g \approx 2.005$ measured by the cross-relaxation R($\spinT{1}{2}$)-Bi($\spinT{8}{13}$) (Ref. \onlinecite{Mortemousque2012}) but its microstructure was unknown. In our measurement, the sample was placed in the JEOL JES-RE3X X band EPR spectrometer.  A small coil placed near the sample within the EPR cavity was used to excite the magnetic resonance.  On the other hand, the X-band ($\approx 9.08$ GHz) irradiation and reflection were used for probing the change in the sample conductivity.  Since the additional coil near the sample could apply an arbitrary microwave frequency, it was possible to reduce the frequency along with the static magnetic field.\cite{Mortemousque2012} 
The second derivative of the reflected X-band intensity with respect to the field modulation was recorded as an SDR signal to reduce the broad cyclotron resonance lines and the background change of the sample conductivity during the magnetic field scan. All the SDR measurements were performed at 16 K.  
%Moreover, the SDR signal also included a strong magnetic field dependent background and recording of the second harmonic of the lock-in detected SDR allowed us to minimize this background. All the measurements were performed at 16 K. 
%The second harmonic of the lock-in detected SDR spectra was recorded at 16 K.

%%%%%%%%%%%%%%%%%%%%%%%%%%%%%%%%%%%%
%%%%%%%%%%%%%%%%%%%%%%%%%%%%%%%%%%%%
	\subsection{Experimental results}
	
\begin{figure}[t]
\includegraphics[width=86mm]{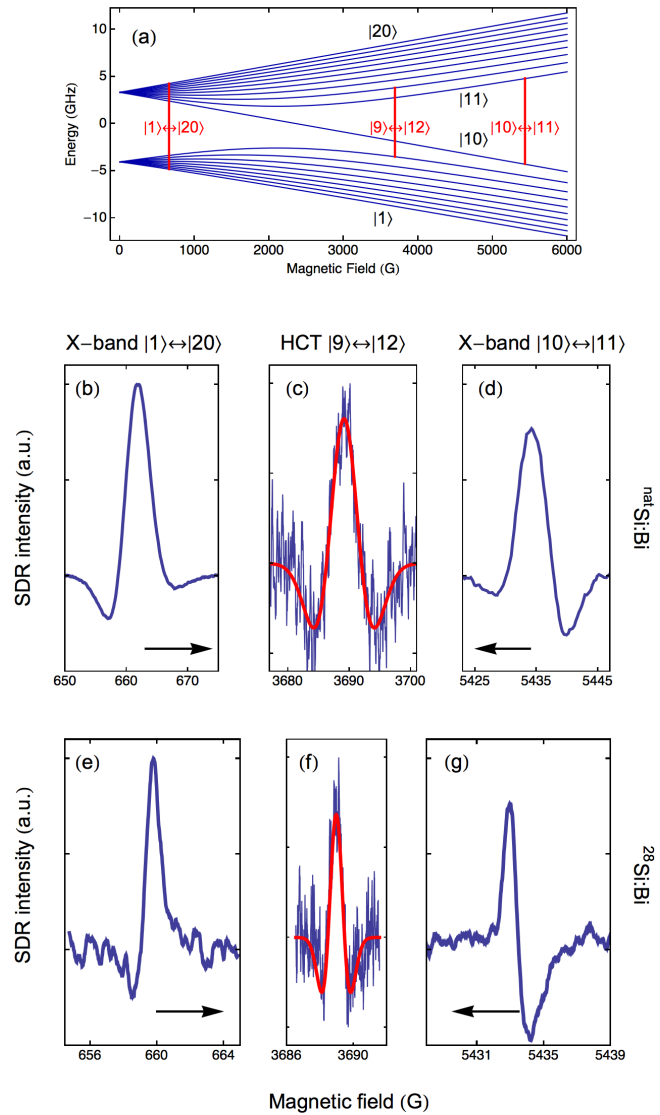}% Here is how to import EPS art
\caption{\label{fig1}(Color online.) (a) Breit-Rabi diagram of the bismuth donor spins. The three vertical red lines correspond to the transitions shown in (b$-$g). cw SDR spectra of \natSi :Bi (b$-$d) and \pureSi :Bi (e$-$g). The FWHM linewidths of the HCT$_{9-12}$  (c) and (f), obtained from the double integration of the fitting Gaussians (red lines), are 6.7 G for \natSi :Bi and is 1.3 G for \pureSi :Bi. Arrows indicate the direction of the asymmetric broadening directions. The signal-to-noise (S/N) ratios for $\spinT{9}{12}$ are worse than the others since the HCT$_{9-12}$ line intensity for Si:Bi is much weaker than the intensities for the X-band $\spinT{1}{20}$ and $\spinT{10}{11}$  lines as will be shown theoretically in Fig. \ref{fig7}.
}
\end{figure}

%%% Hamiltonian
The Bi donor can be modeled by the spin Hamiltonian
\begin{equation}
\label{SpinHamiltonian}
\mathcal{H}_1=g_e \mu_B B_z S_z - g_n \mu_N B_z I_z + h A \bf{S}\cdot\bf{I},
\end{equation}
where $g_e$ and $g_n$ are the donor electron and nuclear $g$-factors, respectively, 
and $A$ the value of the isotropic hyperfine interaction in units of frequency. We label the $i$-th eigenstate in order of increasing energy as $\ket{i}$. The Breit-Rabi diagram of the bismuth donor is shown in Fig. \ref{fig1}(a). The Hamiltonian parameters used are summarized in Table \ref{table1}, together with the ones extracted from the SDR data of this study. 
The sensitivity of the resonant magnetic field to a parameter $p$ for a given resonant frequency $\nu$ is defined as $\delta B_z/\delta p$, which satisfies
\begin{equation}
\label{Eq:SensitivityDef}
\delta \nu = \frac{\partial \nu}{\partial p}\delta p +  \frac{\partial \nu}{\partial B_z}\delta B_z =0
\end{equation}
which leads to $\partial \nu/\partial A = - (\partial \nu/\partial B_z)(\delta B_z / \delta A)$. For $\partial \nu/\partial A$ to be zero, $\delta B_z/\delta A$ must be zero since when $\partial \nu / \partial B_z = 0$, $\delta A = 0$ (Ref. \onlinecite{Wolfowicz2013}) so that $\partial \nu/\partial A$ takes a finite value.

%%%%%%%%%%%%% Table I
\begin{table}[b]
\caption{\label{table1} 
Magnetic resonance parameters of \pureSi :Bi and \natSi :Bi. The fractional changes in $g$-factor and in hyperfine interaction, are calculated as $(g_e^\text{SDR} - g_e^\text{EPR})/g_e^\text{EPR}$ and $(A^\text{SDR} - A^\text{EPR})/A^\text{EPR}$. The values for SDR$^{\text{b,c}}$ i.e., $A^\text{SDR}$ (b, c) and $g_e^\text{SDR}$ (c) are obtained from the fitting of the SDR peak positions by Eq. (1), assuming Gaussian distributed resonance peaks. 
}
\begin{ruledtabular}
\begin{tabular}{lccccc}
\pureSi :Bi &$g_e$&$\Delta g_e/g_e$ & $g_n$ & $A$& $\Delta A/A$\\
&&(ppm)&&(MHz)&(ppm)\\
\hline
EPR\footnotemark[1] & 2.00032 & Ref. & 0.9135 & 1475.17 & Ref.\\
SDR\footnotemark[2] & 2.00036(4) & $+19(22)$ & & 1475.31(7) & $+ 95(50)$\\
SDR\footnotemark[3]& 2.00038(2) & $+29(10)$ &  & 1475.29(7) & $+ 84(50)$\\
\hline
\natSi :Bi &&&&&\\
 \hline
EPR\footnotemark[4]\footnotemark[5] & 2.0003 & Ref. & 0.914 & 1475.4 & Ref.\\
SDR\footnotemark[3] & 2.00049(5) & $+93(25)$ & & 1475.05(17) & $-240$(120)
\end{tabular}
\footnotetext[1]{Wolfowicz \textit{et al.} (Ref. \onlinecite{Wolfowicz2013})}
\footnotetext[2]{$g$-factor and $A$ fitting parameters.}
\footnotetext[3]{$g$-factor determined at HCT$_{9-12}$, $A$ used as fitting parameter}
\footnotetext[4]{Feher (Ref. \onlinecite{Feher1959})}
\footnotetext[5]{Morley \textit{et al.} (Ref. \onlinecite{Morley2010})}
\end{ruledtabular}
\end{table}

%%% SiBi EOWP and spectra explanations
Figures \ref{fig1}(b$-$d) and \ref{fig1}(e$-$g) show cw SDR spectra of \natSi:Bi and \pureSi:Bi.  
The spectra recorded at the Bi donor HCT for $m_I = -7/2$ (7.3043 GHz for \natSi :Bi and 7.3054 GHz for \pureSi :Bi) between the states $\ket{9}$ and $\ket{12}$ [Figs. \ref{fig1}(c, f)] have a symmetric line shape whereas the X-band spectra of the $\spinT{1}{20}$ transition ($m_I = 9/2$) and the $\spinT{10}{11}$ transition ($m_I = -9/2$), shown in Figs. \ref{fig1}(b, e) and in Figs. \ref{fig1}(d, g), respectively, are asymmetric. 
%At the HCT, \txtr{eq. (\ref{Eq:SensitivityDef}) implies that} the resonant field sensitivity to the hyperfine value $\delta B_z / \delta A$ is zero, \txtr{allowing us to} %so that one can 
At the HCT, the resonant field sensitivity to the hyperfine value $\delta B_z / \delta A$ is zero so that one can probe the Si:Bi linewidth and line shape not subject to such electric perturbations. 
%At the HCT, the resonant field sensitivity to the hyperfine value $\delta B_z / \delta A$ \txtr{computed using Eq. (\ref{Eq:SensitivityDef})} is zero, \txtr{allowing us to} 
%probe the Si:Bi linewidth and line shape not subject to such electric perturbations. 
The measured FWHM linewidth of the HCT lines are 6.7 G and 1.3 G for \natSi :Bi  and for \pureSi :Bi, respectively. These values are significantly larger than the measured linewidth at X-band of 4.1 G for \natSi :Bi\cite{Morley2010,George2010} and the theoretical prediction of 0.08 G at the HCT$_{9-12}$ for \pureSi :Bi.\cite{Wolfowicz2012} In contrast, the X-band $\spinT{1}{20}$ transitions are asymmetrically broadened (FWHM linewidths of 7.7 G and 1.6 G) toward high field and the $\spinT{10}{11}$ transitions toward low field (7.0 G and 1.4 G).  $m_I$ dependent asymmetry directions can be described by an (inhomogeneous) distribution of the hyperfine interaction but is inconsistent with any distribution of the Zeeman interaction.  

Let us now discuss whether the experimental conditions we employed are sufficient to achieve the intrinsic linewidth and shape of the Bi donor spin transitions. In the duration of a single measurement, the microwave frequency of the EPR spectrometer drifts typically by $\pm 5$ kHz. The signal generated at 7 GHz by an Agilent 8257D microwave source in series with a 3-W MiniCircuits ZVE-8G+ amplifier exhibits a frequency stability of $\pm 1.5$ kHz for the same duration. These fluctuations in the applied microwaves lead to a maximum line broadening of $\pm 2.5 \times10^{-3}$ G at 9 GHz and $\pm 1 \times10^{-3}$ G at 7 GHz, which is negligibly small compared to the estimated $\gtrsim 10^{-2}$ G precision in magnetic field and its inhomogeneity. 

%
%%%%%%%%%%%%% Figure 1
%
%
%%%%%%%%%%%%% Table II
\begin{table}[b]
\caption{\label{table2} 
The resonant field ($B_z$) sensitivity to the $g$-factor $g_e$ and to the hyperfine $A$ for the Si:Bi transitions shown in Fig. \ref{fig1}. 
 $m_I$ represents the nuclear spin projection of the EPR transitions but is not a good quantum number for all these levels, expect for $\ket{10}$ and $\ket{20}$. 
The calculation was performed using the EPR magnetic resonance parameters shown in Table I.}
\begin{ruledtabular}
\begin{tabular}{lrccc}
Transition &$m_I$ & $\left|\delta B_z/\delta J\right|$ & $\delta B_z/\delta g_e$ &$\delta B_z/\delta A$ \\
(spectrum in Fig. \ref{fig1}) &&[G/MHz]&[$10^3$ G]& [G/MHz] \\
\hline
$\spinT{1}{20}$ (b, e) & $9/2$ & 0.08 & $-0.3$  & $-1.9$  \\
$\spinT{9}{12}$ (c, f) & $-7/2$ & 0.06 & $-1.8$ & $0$ \\
$\spinT{10}{11}$ (d, g) & $-9/2$ & 0.08 & $-2.7$ & $1.4$ \\
\end{tabular}
\end{ruledtabular}
\end{table}

The asymmetric line broadening of the $\spinT{1}{20}$ and $\spinT{10}{11}$ transitions is consistent with a distribution of the donor hyperfine interactions with a long tail toward low hyperfine couplings. From the line shapes of the spectra in Figs. \ref{fig1} (b, d, e and g), the asymmetric part of the line broadening can be estimated roughly 1 G, corresponding to a distribution of the hyperfine constant $A$ toward lower values by 3 MHz. 
We can exclude the distribution in the donor $g$-factor as a cause of this asymmetric broadening, because the sensitivity $\delta B_z / \delta g_e$ is negative for both transitions so that the broadening for both $\spinT{1}{20}$ and $\spinT{10}{11}$ transitions would be in the same direction. 
The spin exchange interaction $J\,\bm{S_{_D}}\cdot\bm{S_{_R}}$ (Ref. \onlinecite{Cox1978}) between the two electrons of the SDR pair can also be ruled out as it would yield a symmetric line broadening for low enough couplings, estimated by Lu \textit{et al.}\cite{Lu2011} to be below 5 MHz for phosphorus donor coupled to a surface dangling bond ($^{31}$P$-$P$_{\text{b}0}$) and below 10 kHz for separations larger than one donor Bohr radius $a_{_B}$.\cite{Suckert2013} The values of $\left|\delta B_z / \delta J\right|$, $\delta B_z / \delta A$ and $\delta B_z / \delta g_e$ corresponding to each spectrum in Fig. \ref{fig1} are summarized in Table \ref{table2}.

%%%%% Strain
Other possible causes for the observed asymmetric broadening 
%change in the donor hyperfine 
would be the strain induced by the implantation damage that was not recovered fully by the post-implantation annealing process.\cite{Kimura1993} 
For shallow donors (P, As, Sb) in silicon, Wilson and Feher\cite{Wilson1961} and Dreher \textit{et al.}\cite{Dreher2011} have shown that uniaxial macroscopic strain decreases the hyperfine interaction mainly through the valley repopulation of the ground-state Bloch function. Recently, Dreher (Ref. \onlinecite{DreherPhD}) has shown that, despite the fact that Bi has a large electron binding energy of 71 meV, the strain decreases its hyperfine interaction in the manner similar to other shallow donors. 
%However, as shown in Table \ref{table1}, the increase in the effective hyperfine of $+ 84$ ppm in \pureSi :Bi was measured  by the present SDR.  
However, the effective hyperfine of the Bi donors in \pureSi, obtained from the peak positions in the SDR spectra, is  $+ 84$ ppm higher than the reported value for EPR measurements.\cite{Wolfowicz2013} 
Thus the macroscopic strain cannot account for the observed positive shift in effective hyperfine interaction. 
In fact, the positive shift suggests that orbitals of the donor and the readout center electrons are coupled and their densities are redistributed. In this study, we thus describe the SDR pair in terms of a model based on the coupling between the electron orbitals of the pair in this study. 

%Also, the amplifier being limited to 8 GHz, we could not perform the spectroscopy of the EOWP of the transitions $\spinT{8}{13}$, $\spinT{7}{14}$, and $\spinT{6}{15}$ occurring, for \pureSi Bi, at 12.57, 23.18, and 72.64 GHz, respectively.

%%%%%%%%%%%%%%%%%%%%%%%%%%%%%%%%%%%%
%%%%%%%%%%%%%%%%%%%%%%%%%%%%%%%%%%%%
%%%%%%%%%%%%%%%%%%%%%%%%%%%%%%%%%%%%	
\section{Calculation of the line shape with the SDR model}

One Bi donor electron and one readout center electron form a spin pair. In section \ref{SDRmodel-wf}, we introduce a theoretical model to describe this electron pair. Then, we evaluate the effect of the readout center on the donor hyperfine properties (section \ref{SDRmodel-hyperfine}) and we discuss the influence of the model parameters on the line shape (section \ref{SDRmodel-parameters}).
% possible explanation to the observed positive shift of the donor $g$-factor (section \ref{SDRmodel-gfactor}).

%%%%%%%%%%%%%%%%%%%%%%%%%%%
%%%%%%%%%%%%%%%%%%%%%%%%%%%
	\subsection{wave function of the donor-readout center pair \label{SDRmodel-wf}}

%%%%%%%%%%%%% Figure 2
\begin{figure}[b]
\includegraphics[width=86mm]{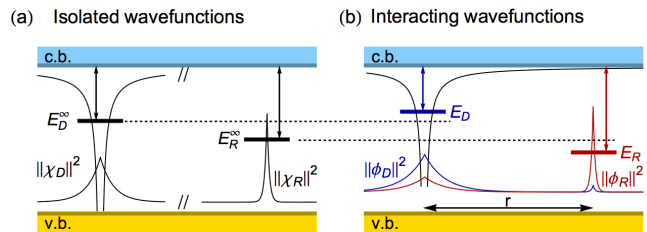}% Here is how to import EPS art
\caption{\label{fig2}(Color online.) 
Energy diagrams of the donor and readout center in the silicon band gap for the isolated states (a) and for the molecular orbitals (b). The corresponding electron densities are also plotted, together with the Coulomb potential of the ionized donor.
}
\end{figure}

The one-electron molecular orbitals corresponding to the neutral donor in the presence of an ionized readout center (D$^0$-R$^+$) and to a neutral readout center close to an ionized donor (D$^+$-R$^0$) are denoted by $\phi_{_D}$ and $\phi_{_R}$, respectively.
In a simplified picture, $\phi_{_D}$ and $\phi_{_R}$ can be expressed as a linear combination of the wave functions of the electron of an isolated donor $\chi_{_D}$ and an isolated center $\chi_{_R}$ so that $\phi_{_D} = a_1 \chi_{_D} + a_2 \chi_{_R}$ and $\phi_{_R} = b_1 \chi_{_D} + b_2 \chi_{_R}$.
The linear coefficients $a_{1,2}$ and $b_{1,2}$ are calculated by applying the variational method to the one-electron Hamiltonian $\mathcal{H}_0 = K^* + V_{_D}^* + V_{_R}^*$ where $K^*$ is the effective kinetic energy of the electron, $V_{_D}^*$ is the screened Coulomb potential of the donor, and $V_{_R}^*$ is the effective potential of the readout center. The difference in energy between these molecular states $\phi_i$ and the isolated states $\chi_i$ is small, even for a small spatial separation. This is due to the significant difference in the two orbitals $\chi_{_D}$ and $\chi_{_R}$. The electron densities $|\chi_{_D}|^2$ and $|\chi_{_R}|^2$ are plotted in Fig. \ref{fig2}(a), and those of the one-electron molecular orbitals $|\phi_{_D}|^2$ and $|\phi_{_R}|^2$ in Fig. \ref{fig2}(b).

Antisymmetrized wave functions of the two-electron system, including the spin part, are constructed using the Slater determinant of the one-electron molecular orbitals:
\begin{subequations}
\begin{align}
\psi_1&=  \phi_{_R} \,\, \phi_{_R} \ket{0,0}\otimes \ket{m_I}\\
\psi_{_+}&= 2^{-1/2}\,\, \big(\phi_{_D} \,\, \phi_{_R} + \phi_{_R} \,\, \phi_{_D}\big)\ket{0,0}\otimes \ket{m_I}\\
\psi_{_-}&=2^{-1/2}\,\, \big(\phi_{_D} \,\, \phi_{_R} - \phi_{_R} \,\, \phi_{_D}\big)\ket{1,m_\sigma}\otimes \ket{m_I} \\
\psi_4&=  \phi_{_D} \,\, \phi_{_D} \ket{0,0}\otimes \ket{m_I}.
\label{Psis0}
\end{align}
\end{subequations}
In the above, the spin states are denoted as $\ket{\sigma, m_\sigma}$ with $\sigma = S_{_D} \pm S_{_R}$ and the orbital products of the $\phi_i$ correspond, from left to right, to the first and the second electrons of the system.
One notices that the spin singlet state $\psi_+$ (triplet $\psi_-$) behaves like a bonding (antibonding) orbital. Note that these states correspond to the charge states D$^+-$R$^-$, D$^0-$R$^0$ ($\sigma$=0), D$^0-$R$^0$ ($\sigma$=1) and D$^--$R$^+$, respectively.

Furthermore, the charge repulsion $1/r_{_{12}}$ can be included. The corrected two-electron molecular orbitals $\Psi_i$ are then written as linear combinations of  $\psi_i$. Thus, the bonding orbital is $\Psi_{_+} = N_+^{-1} \big(\psi_{_+} + c_1 \psi_1+ c_4 \psi_4 \big)$, where the coefficients are $c_1 = \frac{\left<\psi_1 | 1/r_{_{12}} | \psi_{_+} \right>}{E_+-E_1}$ and $c_4 = \frac{\left<\psi_4 | 1/r_{_{12}} | \psi_{_+} \right>}{E_+-E_4}$. 
%The contribution from the spin triplet state $\psi_{_-}$ vanishes as the perturbation $1/r_{_{12}}$ does not mix the different spin states and the orbital part of the matrix element $\matrixel{\psi_{_-}}{1/r_{_{12}}}{\psi_{_+}}$ vanishes. 
Then, assuming a negatively charged donor (D$^-$) with an energy $\approx E_4 \gg E_+, E_1$, the coefficient $c_4$ tends to zero and the contribution of $\psi_4$ to $\Psi_{_+}$ can be neglected. It follows that
\begin{equation}
\Psi_{_+} = N_+^{-1}\Big(\psi_+ + c_1 \psi_1 \Big).
\end{equation}
In the coefficient $c_1$, the term $\left<\psi_1 | 1/r_{_{12}} | \psi_{_+} \right> $ can be approximated as $\approx \sqrt{2}\, \theta\, \correnergy$ where $\correnergy$ is the two-electron correlation energy taken as the Coulomb repulsion of the electrons in the R$^-$ state. 
In this model, $\correnergy$ is included in the parameter $\corrparam =\correnergy / (E_+ - E_1) \approx \correnergy / (E_+ - E_{_R-})$. 
On the other hand, the antibonding spin triplet state $\psi_{_-}$ does not mix with either of the spin singlet states $\psi_{1,+,4}$, i. e., we have $\Psi_{_-} = \psi_{_-}$.

%%%%%%%%%%%%%%%%%%%%%%%%%%%
%%%%%%%%%%%%%%%%%%%%%%%%%%%
	\subsection{Change in hyperfine interaction \label{SDRmodel-hyperfine}}

%%%%%%%%%%%%% Figure 3
\begin{figure}[b]
\includegraphics[width=76mm]{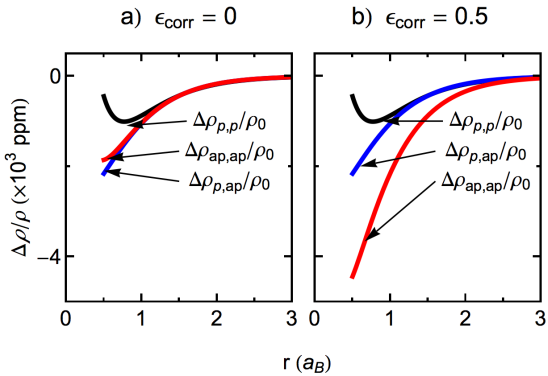}% Here is how to import EPS art
\caption{\label{fig3}(Color online.) 
%Fractional changes of the hyperfine interactions normalized by the spin projections ($A_{i,j} = -2/3\,\mu_0\, \mu_B\, \mu_N\, \rho_{i,j}(r_{\text{Bi}})$) plotted as a function of the separation $r$ between the donor and the readout center in the unit of the Bohr radius of the bismuth donor.
Fractional changes in the electron density $\rho$ for three different electron spins configurations plotted as a function of the separation $r$ between the donor and the readout center in units of $a_{_B}$. $\rho_0$ corresponds to the isolated bismuth donor. 
A typical fractional change of $- 2\times10^{3}$ ppm corresponds to a change of the Bi hyperfine interaction of $- 3$ MHz. 
%The effect of the $g$-factor distribution is not included in this plot and only the density of electron at the bismuth nucleus is considered.
}
\end{figure}

The Fermi hyperfine interaction for the two electrons is
\begin{equation}
\label{hyperfineformula}
\mathcal{H}_{\text{hyp}} = -\frac{2}{3}\,\mu_0 \,\,\bm{\mu}_{_\text{Bi}} \cdot \sum_{i=1}^2 \bm{\mu}_i \rho_i(\bm{r_{_\text{Bi}}}),
\end{equation}
where $ \rho_i(\bm{r_{_\text{Bi}}})$ is the one-electron density at the bismuth nucleus. The electron magnetic dipolar moment $\bm{\mu}_i$ depends on the electron orbital function. As the two-electron orbitals can be expressed as functions of $\chi_{_D}$ and $\chi_{_R}$, only two operators $\bm{\mu}_{_D} = - g_{_D} \,\mu_B \bm{S}_{_D}$ and $\bm{\mu}_{_R} = - g_{_R} \, \mu_B \bm{S}_{_R}$ are relevant, where $g_{_D}$ and $g_{_R}$ are the $g$-factors of the isolated donor and readout center electrons, respectively. 
In order to simulate the SDR line shape, we only consider the change in the electron distribution while assuming the $g$-factor of the isolated centers. However, due to the confined nature of the readout center, only the $\chi_{_D}$ component has a significant electron density at the bismuth nucleus. Then, in the rest of this section, the subscript of $\rho_{_D}$ is dropped.

Now, if one considers the hyperfine interaction $A_{m_{_R}}$ for a given spin projection $m_{_R}$ of the readout center, one finds that
\begin{equation}
\matrixel{m_{_R}=1/2}{\mathcal{H}_\text{hyp}}{m_{_R}=1/2} =
\begin{pmatrix}
\mathcal{A}_{p,p} &  \mathcal{A}_{p,ap}\\
\mathcal{A}_{ap,p} &  \mathcal{A}_{ap,ap}
\end{pmatrix}
\end{equation}
\begin{equation}
\matrixel{m_{_R}=-1/2}{\mathcal{H}_\text{hyp}}{m_{_R}=-1/2} =
\begin{pmatrix} 
\mathcal{A}_{ap,ap}&\mathcal{A}_{p,ap}\\
\mathcal{A}_{ap,p} &  \mathcal{A}_{p,p}
\end{pmatrix}
\end{equation}
where each $\mathcal{A}_{j,k}$ on the right-hand side is a block matrix of dimension $2I+1$, calculated using the electron density $\rho_{j,k}$ with subscripts indicating the parallel and antiparallel electron spin configurations: $\mathcal{A}_{p,p} = \matrixel{\Psi_-}{\mathcal{H}_{\text{hyp}}}{\Psi_-}$, $\mathcal{A}_{p,ap} = \matrixel{\Psi_-}{\mathcal{H}_{\text{hyp}}}{(\Psi_--\Psi_+)/\sqrt{2}} = \matrixel{\Psi_-}{\mathcal{H}_{\text{hyp}}}{(\Psi_-+\Psi_+)/\sqrt{2}}$ and $\mathcal{A}_{ap,ap} = \matrixel{(\Psi_--\Psi_+)/\sqrt{2}}{\mathcal{H}_{\text{hyp}}}{(\Psi_--\Psi_+)/\sqrt{2}} = \matrixel{(\Psi_-+\Psi_+)/\sqrt{2}}{\mathcal{H}_{\text{hyp}}}{(\Psi_-+\Psi_+)/\sqrt{2}}$. 
On the other hand, the off-diagonal blocks $\matrixel{m_{_R}'}{\mathcal{H}_\text{hyp}}{m_{_R}}$ for $m_{_R}' \neq m_{_R}$ give a contribution only at the second and higher orders, which are neglected in this model. 
The simulation of the fractional change in the electron density at the donor nucleus $\Delta \rho /\rho_0$ was performed using a single exponential envelope function characterized by the Bohr radius $a_{_B}=8.1$ \AA$\,$ for the Bi donor electron and a Dirac function for the readout center.  $\Delta \rho /\rho_0$ is plotted in Fig. \ref{fig3} for a readout center energy of $-0.55$ eV, and repulsion energy parameters $\corrparam = 0$ (a) and $\corrparam = 0.5$ (b). One notices that a large repulsion energy parameter decreases the hyperfine interaction for the electron spin pair in the triplet configuration.

%%%%%%%%%%%%%%%%%%%%%%%%%%%
%%%%%%%%%%%%%%%%%%%%%%%%%%%
	\subsection{SDR model parameters \label{SDRmodel-parameters}}
The present model contains three physical parameters for a given donor in silicon: the concentration of readout centers $N_{_R}$, and two parameters $E_{_R}$ and $\corrparam$ related to the energy levels of the readout center.
In order to discuss the effect of the model parameters on the spectral line shapes, it is required to know how much each SDR pair contributes to the detected SDR signal as a function of the pair separation distance.

%%%%%%%%%%%%%%%%%%%%%%%%%%%%%%%%%%%%
%\subsubsection{Concentration of the readout centers $N_{_R}$ \label{SDRmodel-parameters-Nr}}
	
%%%%%%%%%%%% Figure 4
\begin{figure}[t]
\includegraphics[width=86mm]{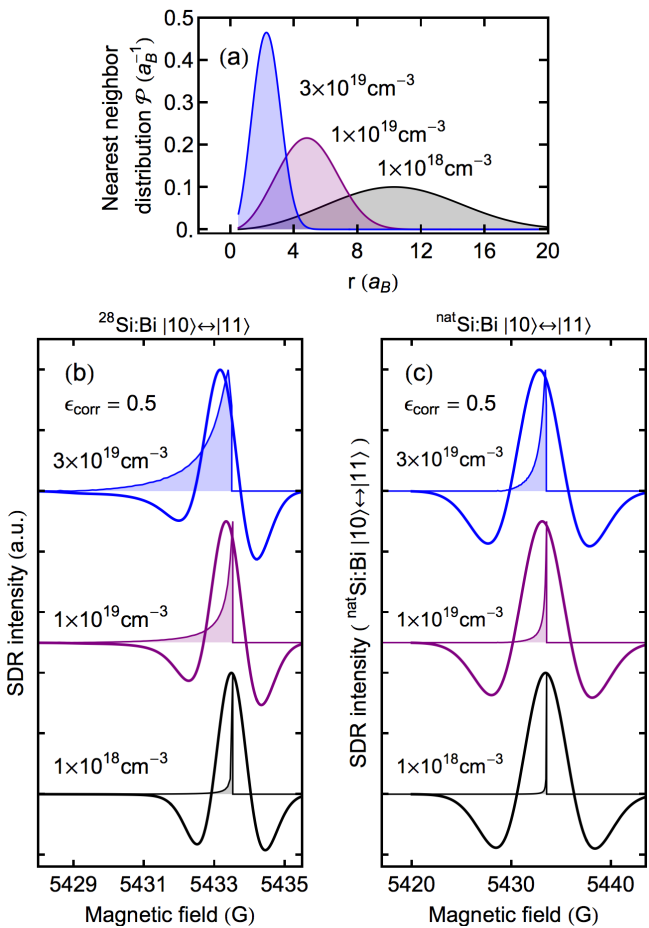}% Here is how to import EPS art
\caption{\label{fig4}(Color online.) (a) Distributions of the separation $r$ between the donor and its nearest readout center for various  concentrations $N_{_R}$.  
%The radius $r$ and the probability density function $\mathcal{P}$ are expressed in unit of the Bi Bohr radius $a_0$. 
(b, c) Simulated distributions of the $\spinT{10}{11}$ transition taking into account only the SDR pair distribution (thin filled lines) and its convolution with the second derivative of a Gaussian (thick lines) for \pureSi :Bi (b) and \natSi :Bi (c). 
The donor and readout center pair with the small enough separation $r$ have strong interaction and thus contribute to the low-field tail in the distribution of resonant magnetic fields (thin solid lines).  
The FWHMs of the HCT$_{9-12}$ lines are 6.7 G for \natSi :Bi (c) and 1.3 G for \pureSi :Bi (f), each of which is obtained from the width parameter of the 2nd derivative of a Gaussian function fitted to the HCT line (red curves). 
%The Gaussian FWHM linewidths are 1.35 G for \pureSi :Bi and 6.7 G for \natSi :Bi (measured at HCT$_{9-12}$ for both materials and corrected for the change in $\delta B_z / \delta g_e$, see text). 
Due to such linewidth difference, the shift of the peak position to low field is much larger in \natSi :Bi (larger field scale), while the degree of line shape asymmetry is more apparent in \pureSi :Bi, as $N_{_R}$ is increased.
%In the convoluted simulation with $N_R$ being increased, a shift of the peak position to low field is more apparent in \natSi :Bi while the line shape asymmetry (deformation) is more significant in \pureSi :Bi, due to different Gaussian linewidths.
}
\end{figure}

Among all the readout centers interacting with a donor, we assume that the closest one exclusively forms the most efficient recombination pair.  Then,  in the ensemble measurement, each donor has a different separation $r$ to the nearest readout center and, therefore, a different recombination time in the anti-parallel spin configuration, $\tau_{ap}$.  However $\tau_{ap}$ is much shorter than the pair creation time $\tau_{ec}$, i. e.,  $\tau_{ap} \ll \tau_{ec} \ll \tau_{p}$,  the signal intensity from a single D-R pair is determined by the electron capture time $\tau_{ec}$ 
%rather than by the distribution in $\tau_{ap}$ in our continuous steady-state measurements. 
and thus independent of $r$ in cw SDR measurements. 
Then, the total intensity from an ensemble of D-R pairs should be determined directly by the distribution function of the D-R separation $r$.  
%However, the signal intensity increases with the number of pairs, therefore, the pair formation is the crucial element of the SDR measurement.  
Here we identify the concentration of the pair having the pair separation $r$ as follows. 

First, we assume that the SDR intensity is proportional to the probability $\mathcal{P}(r)\,dr$ of a Bi donor to find the nearest readout center at a distance between $r$ and $r+dr$. %in a shell of volume $4\pi r^2$d$r$.
%This distribution has been calculated in an early work by Paul Hertz in 1909. 
%For a large concentration $N_{_R}$ of readout centers, i. e.,  the concentration of bismuth donors is much less\cite{Marsh1968,Baron1969,deSouza1993,Weis2012} than $N_{_R}$, one obtains
This distribution can be written as\cite{Hertz1909}
\begin{equation}
\label{HertzDistribution}
\mathcal{P}=\frac{3}{\left<r_{_{RR}}\right>}\left(\frac{r}{\left<r_{_{RR}}\right>}\right)^2 \exp \left(-\frac{r^3}{\left<r_{_{RR}}\right>^3}\right)
\end{equation}
where $\left<r_{_{RR}}\right> = (3V / 4 \pi N_{_R})^{1/3}$ is the average distance between the readout center and its nearest neighbor.
Such distributions are plotted in Fig. \ref{fig4}(a), as a function of $r$ in the unit of $a_{_B}$, for three different concentrations $N_{_R}$ of the readout centers. 
%On the other hand, in \natSi :Bi, the peak line width is limited by the inhomogeneous broadening due the presence of 4.7$\%$ of $^{29}$Si whereas in \pureSi :Bi, it is limited by the dipolar coupling to other bismuth donor ($\sim 1\times 10^{18}$ cm$^{-3}$) and readout center electron spins ($\sim 1\times 10^{19}$ cm$^{-3}$). Therefore, the SDR data can be described using a Gaussian line shape for both materials. 
By combining Eq. (\ref{HertzDistribution}) with the dependence of the hyperfine $A$ on the D-R separation $r$ obtained in section \ref{SDRmodel-hyperfine}, the distribution in resonant magnetic field for the transition $\spinT{10}{11}$ is calculated and shown by thin curves in Figs. \ref{fig4} (b, c). Since the peak for each $r$ should be accompanied by a symmetric broadening due to inhomogeneous distribution of $^{29}$Si nuclear spins in \natSi :Bi and of other Bi-donor and readout-center electron spins in \pureSi :Bi [as observed in Figs. 1(c) and 1(f)], the thin curves are convoluted with the second derivative of a Gaussian function to simulate the SDR spectra. The simulated spectra are shown as the thick curves in the same figures. 

%%%%%%%%%%%%%%%%%%%%%%%%%%%%%%%%%%%%
%\subsubsection{Energy levels of the readout center $E_{_R}$ and $\corrparam$\label{SDRmodel-parameters-Er}}

%%%%%%%%%%%% Figure 5
\begin{figure}[b]
\includegraphics[width=86mm]{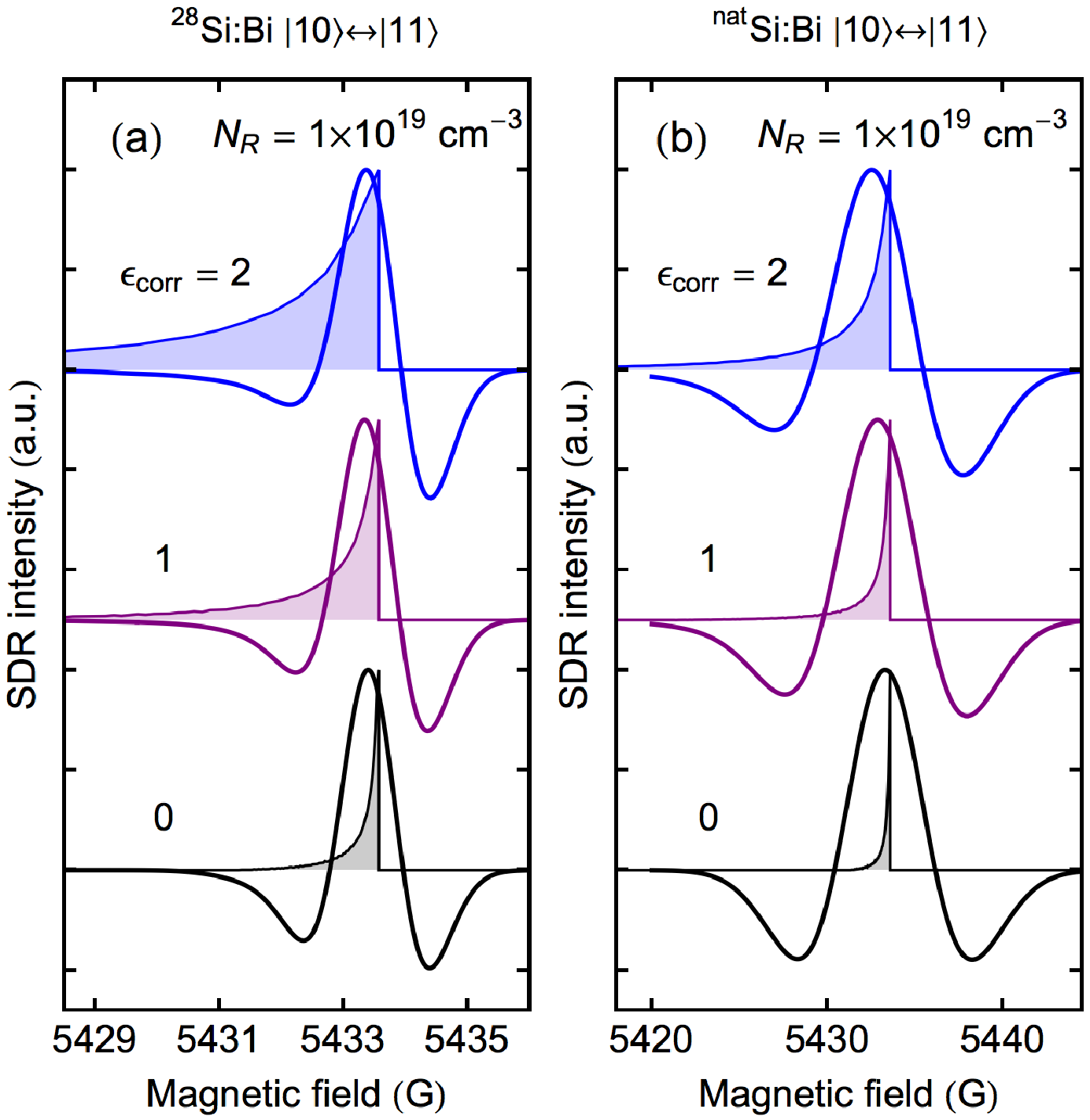}% Here is how to import EPS art
\caption{\label{fig5}(Color online.) Simulated distribution of the $\spinT{10}{11}$ transition in \pureSi :Bi (a) and \natSi :Bi (b) taking into account only the SDR pair distribution (thin lines) and its convolution with the second derivative of a Gaussian (thick lines) for various two-electron correlation parameters $\corrparam$. A larger $\corrparam$ decreases the hyperfine interaction more in the anti-parallel spin pair configuration, which in turns broadens the distribution of the resonant magnetic field toward low field. The same procedure as for Fig. \ref{fig4} was used for these simulations.}
\end{figure}

%The readout center energy plays a role as a mixing parameter in the determination of the one-electron wave functions. Larger the ionization energy $E_{_R}$ corresponding to the $\phi_{_R}$ state in Fig. \ref{fig2}(b), higher the degree of mixing of the atomic orbitals leading decrease in the donor hyperfine. 

The mixing of atomic orbitals in the present model  is assumed to be driven by the long range Coulomb potential of the ionized donor, and the readout center energy $E_{_R}$ is set at $-0.55$ eV from the silicon conduction band. The remaining parameter of this model is the two-electron correlation parameter $\corrparam$ defined in section \ref{SDRmodel-wf}. It characterizes the mixing of the two-electron molecular orbitals in the spin singlet configuration $\Psi_+$. The dependence on $\corrparam$ of the resonant magnetic field is plotted in Fig. \ref{fig5} for the $\spinT{10}{11}$ transition.

%%%%%%%%%%%%%%%%%%%%%%%%%%%%%%%%%%%%
%%%%%%%%%%%%%%%%%%%%%%%%%%%%%%%%%%%%
%%%%%%%%%%%%%%%%%%%%%%%%%%%%%%%%%%%%
\section{Comparison of experimental results and simulations}

%%%%%%%%%%%%%%%%%%%%%%%%%%%%%%%%%%%%
%%%%%%%%%%%%%%%%%%%%%%%%%%%%%%%%%%%%
	\subsection{Line position\label{Discussions-Position}}
At low magnetic field that we employed, the line positions are determined by the two parameters, $g_{_D}$ and $A$. In section III, we have shown the dependence of the resonant field on the electron density at the donor nucleus, $\rho$, through the hyperfine interaction. The donor electron $g$-factor further influences the line positions through both the Zeeman and the hyperfine [Eq. (\ref{hyperfineformula})] interactions. 
Because the resonant magnetic field of the HCT$_{9-12}$ is robust against fluctuations in hyperfine $A$, it allows a precise determination of the $g$-factor of the donor electron. 
We measured an effective shift in the donor electron $g$-factor of $+29$ ppm in \pureSi :Bi (see Table \ref{table1}), which can be qualitatively explained by the second order perturbation theory as follows. 
For a donor electron non interacting with any readout center, the deviation $\delta g_{_D}^\infty$ from the free electron $g$-factor resulting from the spin-orbit coupling is given by:
\begin{equation}
\delta g_{_D}^\infty \mu_B B_z S_z = \sum_{n\neq\chi_{_D}^\infty}\frac{\left< \chi_{_D}^\infty \left| \mathcal{H}_2 \right| n \right> \left< n \left| \mathcal{H}_2 \right| \chi_{_D}^\infty \right>}{E_{\chi_{_D}^\infty} - E_n}
\end{equation}
where 
$E_n$ are eigenvalues of the Hamiltonian $\mathcal{H}_0$ and 
$\mathcal{H}_2 = g_{\text{fe}} \mu_B \, \bm{S} \cdot \bm{B} - \lambda \, \bm{S}\cdot\bm{l} + \mu_B \, \bm{l}\cdot\bm{B}$ 
with $g_\text{fe}$ the free electron $g$-factor, and $\lambda$  the spin-orbit coupling parameter. 
Here the electron ground state $\ket{\chi_{_D}^\infty}$ is an eigenstate of $\mathcal{H}_0$, neglecting the readout center potential $V_{_R}^*$. However, as shown in section \ref{SDRmodel-wf}, the electron wave function is modified due to the presence of the readout center. Therefore, the $g$-factor correction of the donor electron in an SDR pair is $\delta g_{_D} \approx  a_1^2 \delta g_{_D}^\infty + a_2^2\delta g_{_R}^\infty$ where $\delta g_{_D}^\infty$ and $\delta g_{_R}^\infty$ are the spin-orbit corrections of the isolated donor and readout center, respectively, and $a_{1,2}$ are defined in section \ref{SDRmodel-wf}. 
Since the $g$-factor of the isolated readout center $g_{_R}^\infty = 2.005(3)$ (Ref. \onlinecite{Mortemousque2012}) is larger than the $g$-factor of the isolated donor $g_{_D}^\infty = 2.00032$ (see Table \ref{table1}), the weighted average $g_{_D}$ must satisfy $g_{_D}^\infty < g_{_D} < g_{_R}^\infty$.
This qualitatively explains the larger effective $g$-factor of the donor in an SDR pair $g_{_D} = 2.00038(2)$.
Moreover, since the hyperfine interaction is proportional to the donor $g$-factor [Eq. (\ref{hyperfineformula})], the positive change of $+29$ ppm in $g$-factor measured in the SDR spectroscopy of \pureSi :Bi can be partly accounted for by the increase in effective hyperfine interaction of $+84$ ppm. On the other hand, the linewidth of a transition in \natSi :Bi is much larger than in \pureSi :Bi due to the inhomogeneous hyperfine interaction with the $^{29}$Si nuclear spins. Therefore, the line position where the SDR intensity has a maximum,  
%the minimum in microwave absorption,
is shifted toward the mean of the resonant field distribution, away from its maximum (see Fig. \ref{fig4} and \ref{fig5}). Thus, the decrease in effective hyperfine of $-240$ ppm for \natSi :Bi is attributed to a combination of the line asymmetry from the distribution in resonant magnetic field and of the broad linewidth from the inhomogeneous broadening.

The excitation frequency of the HCT$_{9-12}$ (Fig. \ref{fig1}) has been determined using the reference values of the donor electron $g$-factor $g_e^\text{EPR}$ (Table \ref{table1}). However, the $g_e^\text{SDR}$ measured by SDR spectroscopy is different from $g_e^\text{EPR}$. The resulting deviations in resonant field HCT$^{\text{EPR}}_{9-12}$ $-$ HCT$^{\text{SDR}}_{9-12}$  are $+ 0.11$ G for \pureSi :Bi and $ + 0.35$ G for \natSi :Bi. As a consequence, the spectra of Figs. \ref{fig1}(b) and \ref{fig1}(e) are not exactly at the HCT$^{\text{SDR}}_{9-12}$, and the sensitivity $\delta B_z / \delta A (B_z=B^{\text{EPR}}_{\text{HCT}})$ is finite: $+3\times 10^{-8} $ G/MHz for \pureSi :Bi and $+8 \times 10^{-8} $ G/MHz for \natSi :Bi. Nevertheless, the line broadening due to these finite sensitivities is much smaller than the magnetic field inhomogeneity and cannot be detected.

%%%%%%%%%%%%%%%%%%%%%%%%%%%%%%%%%%%%
%%%%%%%%%%%%%%%%%%%%%%%%%%%%%%%%%%%%
	\subsection{Line shape\label{Discussions-Shape}}
	
\begin{figure}[b]
\includegraphics[width=86mm]{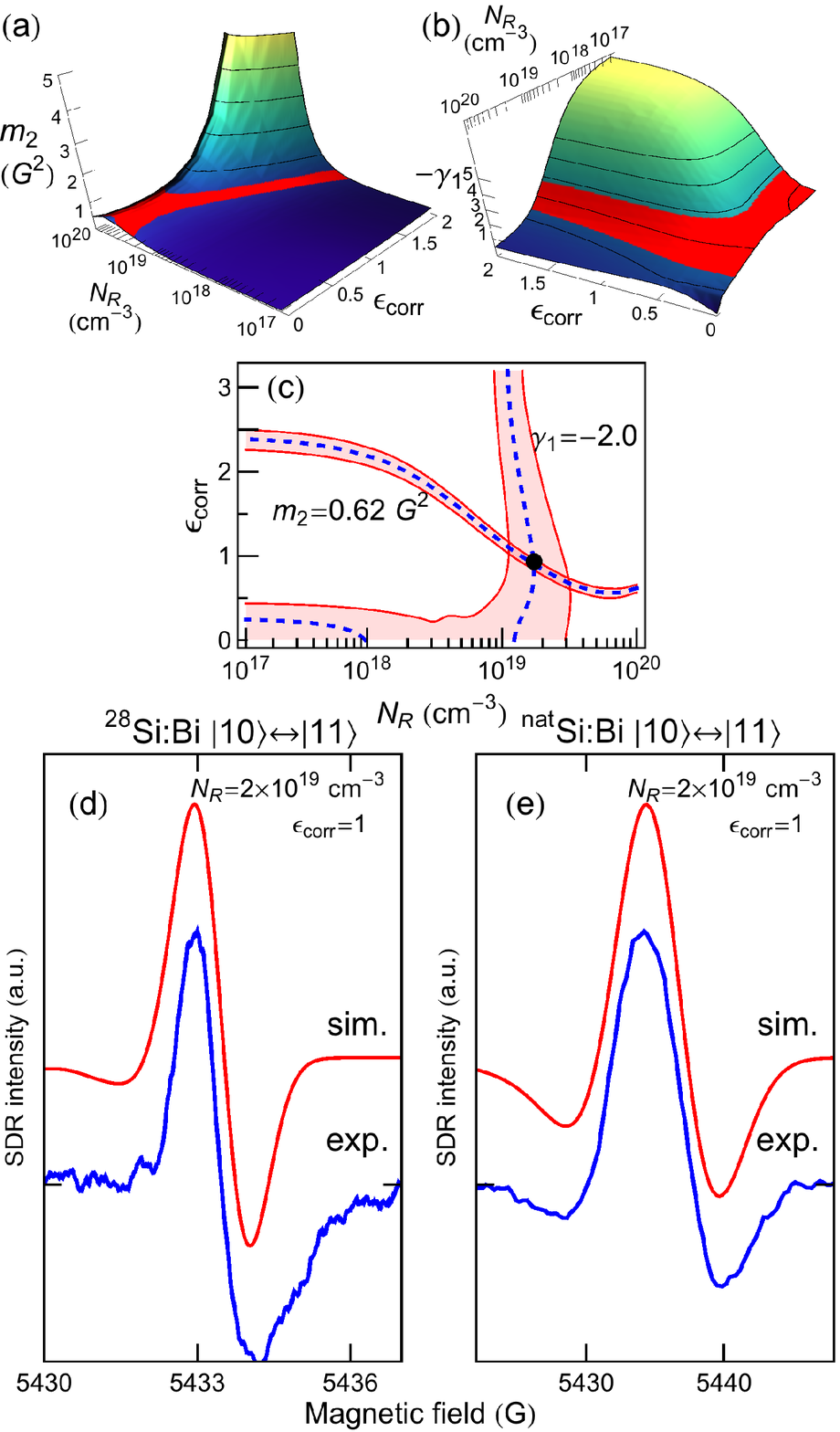}% Here is how to import EPS art
\caption{\label{fig6}(Color online.) Second (a) and third (b) standardized moments ($m_2$ and $\gamma_1$) of the simulated fractional change in the photoconductivity for the transition $\spinT{10}{11}$. The simulation was performed for \pureSi :Bi using the same 1.35 G linewidth as for Fig. \ref{fig4}(b) and \ref{fig5}(a). 
The red regions in the both plots represent the experimental values of $m_2$ and $\gamma_1$ and their uncertainty. These two regions are superposed in (c).  The intersection of the $m_2=0.62$ G$^2$ and $\gamma_1=2.0$ is shown by a filled circle. 
The simulated line shapes for the pinpointed parameters in (c) of \natSi :Bi and \pureSi :Bi are shown in (d) and (e) (red lines), and compared to the experimental data (blue lines).}
\end{figure}

The experimental and simulated line shapes can be quantitatively compared in terms of moments %f moments analyzed by comparing their moments
 $m_n$ defined as:
\begin{equation}
\label{discretemoment}
%m_n  = \Delta B_z\sum_{i=1}^N \big(B_{z,i} - \left<B_z\right>\big)^n\, \mathcal{I}(B_{z,i})
m_n = \int{ (B-\left<B\right>)^n\, \mathcal{I}(B)\, dB}
\end{equation}
where $\mathcal{I}$ is the normalized signal intensity and $\left<B\right>$ is the mean field for this spectrum. 
The degree of broadening and asymmetry can be represented by the variance $m_2$ and skewness $\gamma_1=m_3/m_2^{3/2}$. 
%The odd moments vanish for symmetric lines, and the degree of asymmetry can be represented by the skewness $\gamma_1=m_3/m_2^{3/2}$ of the line distribution. 
%Therefore, the signal, recorded as the second derivative of the sample photoconductivity, has to be integrated twice before the evaluation of the moments. These integrations make the evaluation of the line variance $m_2$ and skewness $\gamma_1$ uncertain because of the large signal background.
The simulated values of $m_2$ and $\gamma_1$ for \pureSi :Bi are plotted as functions of $N_{_R}$ and $\corrparam$ in Figs. \ref{fig6}(a) and \ref{fig6}(b), respectively. 
The experimental variance and skewness are, $m_2 = 0.62(5)$ G$^2$ and $\gamma_1 = -2.0(4)$ for the spin transition $\spinT{10}{11}$ in \pureSi :Bi.
These are represented by the red surfaces in  Figs. \ref{fig6}(a) and \ref{fig6}(b). 
The experimental uncertainties come mainly from a large background after the double-integration of the SDR signal, which is recorded as the second derivative of the sample photoconductivity, necessary for the intensity in Eq. (10) to evaluate the moments. The intersection in Fig. 6(c) represents the corresponding values for the correlation parameter and the readout center concentration: $\corrparam$ = 1 and $N_{_R} =2\times10^{19}$ cm$^{-3}$.  
Such a high readout center concentration is consistent with the high damage cross-section for energetic bismuth ions and the limited recovery of the crystallinity by the annealing process. The two-electron correlation parameter $\corrparam$ = 1 obtained in this study is equal to the one estimated for $^{31}$P-P$_{\text{b0}}$ ($\corrparam \approx 1.0$, Ref. \onlinecite{Poindexter1984}), which confirms the localized wave function of the readout center.

 %The admissible values of $\corrparam$ and $N_{_R}$  intersect for $\corrparam = 0.4$ and $N_{_R} =9\times10^{18}$ cm$^{-3}$ (Fig. \ref{fig6}(c)). Such a high readout center concentration is consistent with the high damage cross-section for energetic bismuth ions and the limited recovery of the crystallinity by the annealing process. The two-electron correlation parameter $\corrparam=0.4$ obtained in this study is of the same order as $\corrparam \approx 1$ obtained for $^{31}$P-P$_{\text{b0}}$ (Ref. \onlinecite{Poindexter1984}).

%%%%%%%%%%%% Figure 6

For these numerical simulations, we used the experimental linewidth of \pureSi :Bi (1.3 G) measured at HCT$_{9-12}$. This rather large linewidth can be explained by the dipole-dipole interaction of the donor and the readout center electron spins for a concentration $N_{_R}\approx 5\times10^{18}$ cm$^{-3}$.
Moreover, one can expect a spectral line broadening due to the distribution in the donor electron $g$-factor. Assuming that this distribution covers a range of $\pm 93$ ppm around $g_{_D}=2.00049$ for \natSi :Bi (see Table \ref{table1}), the broadening in the line FWHM, induced by the finite sensitivity $\left|\delta B_z / \delta g_e \right|$ (see Table II) at the HCT, should be $+ 0.3$ G. As a consequence, the distribution in the donor electron $g$-factor is negligible for \natSi :Bi and the FWHM linewidth of the Gaussian for the transition $\spinT{10}{11}$ is 5.7 G. 
For \pureSi :Bi however, even a smaller distribution of $+29$ ppm in $g$-factor is responsible for 0.1 G linewidth broadening (more than 10\% of the linewidth measured at the HCT$_{9-12}$). The 0.1 G contribution of the $g$-factor distribution to the linewidth is multiplied by the sensitivity ratio $\left(\delta B_z/\delta g_e\right)_{\left|10\right> \leftrightarrow \left|11\right>}/\left(\delta B_z/\delta g_e\right)_\text{HCT}=1.5$. Thus,  the Gaussian linewidth to be used in the simulations for \pureSi :Bi is 1.35 G. It can be noted that for close pairs ($r < 1 \,a_{_B}$), the strong exchange interaction\cite{Suckert2013} can be neglected since the corresponding SDR intensity for $N_{_R} = 2\times10^{19}$ cm$^{-3}$ is below 0.1 \% of the total SDR intensity.
The above mentioned linewidths together with the $N_{_R}$ and $\corrparam$ parameters calculated for \pureSi :Bi lead to the simulated spectra shown in Fig. \ref{fig6}(d) for \pureSi :Bi and (e) for \natSi :Bi. The experimental spectra are also shown below the simulations. The line shapes of the transition $\spinT{10}{11}$ for both \pureSi :Bi and \natSi :Bi samples are well reproduced. This demonstrates the validity of the presented molecular model for the SDR detection of donors for a wide range of host isotope composition.

%The presented model can be further improved by taking into account the hybridization of the donor wave functions (1sA$_1$) with excited states. This hybridization may, in the same way as for strain, mix the donor ground state with hyperfineless states such as 1s$E$ and 1s$T_2$ states, and reduce the donor hyperfine through the valley repopulation process. Also, the undulations of the Bloch functions may modulate the coupling with the readout centre depending on their respective location in the host crystal. In addition, the readout centre wave function could be described with a larger set of functions basis. On the other hand, other experiments such as deep level transient spectroscopy or time-resolved SDR-MR may give more information about the readout centre energy levels,\cite{Poindexter1984}and dissociation and recombination rates.

Before concluding this section, we would like to point out the work of Morishita \textit{et al.}\cite{Morishita2011} in which the spectroscopy of \pureSi :P was performed using low-field electrically detected magnetic resonance (LFEDMR), a technique similar to SDR. In this work, the authors compared the linewidth of \pureSi :P probed by LFEDMR at 160 MHz and by EPR at 9 GHz. No difference in the linewidth ($0.1$ G) for the $\spinT{2}{3}$ transition was observed and the authors concluded that the interaction of the phosphorus donor with the readout center is strong enough to allow the recombination process, but weak enough not to alter the transition linewidth. Yet, the hyperfine structure of the phosphorus donor is only 117 MHz so that its maximum change due to the interaction with the readout center is $\sim 13$ times smaller for phosphorus than for bismuth. Moreover, the small phosphorus nuclear spin $I=1/2$ makes the sensitivity $\delta B_z / \delta A$ relatively small: $-0.10$ G/MHz at 160 MHz for the $\spinT{2}{3}$. Thus, the effect of the phosphorus donor interaction with its readout center on the magnetic resonance is below the detection limit and the conclusions of Morishita do not contradict the present analysis.

%Before concluding, we would like to point out the work of Morishita, \textit{et al.}\cite{Morishita2011} in which the linewidth of phosphorus in $^\text{nat}$Si and $^{28}$Si at low- and high-field are compared. No significant difference in the linewidth between the $\spinT{2}{3}$ transition at 160 MHz and X-band was reported. However, for this two points, the values of $\delta B_z/\delta A$ are low and of the same order, that is to say -0.10 and -0.18 G/MHz, respectively. Together with the weak hyperfine interaction for phosphorus in silicon of 117.53 MHz, the 1000 ppm ranged hyperfine distribution causes an asymmetric broadening of $\sim 0.01$ G at low-field, and 0.02 G at X-band, which is much lower than the observed width of 0.1 G.

%%%%%%%%%%%%%%%%%%%%%%%%%%%%%%%%%%%%
%%%%%%%%%%%%%%%%%%%%%%%%%%%%%%%%%%%%
%%%%%%%%%%%%%%%%%%%%%%%%%%%%%%%%%%%%
\section{Hyperfine clock transitions for other group-V donors in silicon}

%%%%%%%%%%%%% Figure 7
\begin{figure}[t]
\includegraphics[width=86mm]{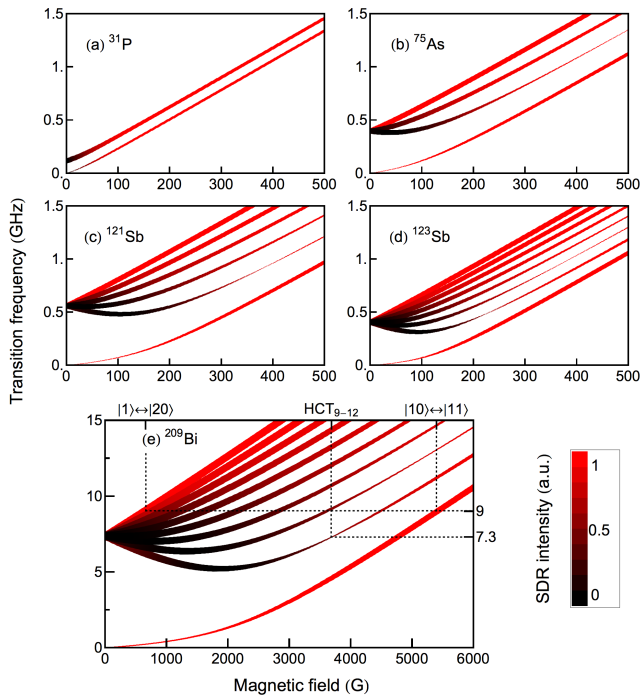}% Here is how to import EPS art
\caption{\label{fig7}(Color online.) EPR transition frequencies of group-V donors in silicon, (a) $^{31}$P, (b) $^{75}$As, (c) $^{121}$Sb, (d) $^{123}$Sb and (e) $^{209}$Bi. The line thickness is proportional to the absolute $|\partial \nu/\partial A|$ value. The color represents the expected SDR intensity for a conventional continuous wave measurement. Our model shows that the SDR intensity of the HCT$_{9-12}$ line for Si:Bi is much weaker than the intensities for the X-band $\spinT{1}{20}$ and $\spinT{10}{11}$ lines. This was observed experimentally in Fig. \ref{fig1}. The frequency scale for $^{209}$Bi is ten times larger than the others.} 
\end{figure}

There is no HCT in the EPR transitions of phosphorus donors in silicon. Other group-V donors have $I -1/2$ HCT. At such points, as discussed in section \ref{Discussions-Position}, the contribution of the $g$-factor distribution to the linewidth can be evaluated knowing the intrinsic EPR linewidth and extrapolated for an arbitrary transition. In fact, since the broadening due to the distribution in hyperfine (electron density at the donor nucleus) scales with $\left| \partial \nu / \partial A \right|$, the contribution to the linewidth calculated in this paper can be extrapolated for any points. The values of $\left| \partial \nu / \partial A\right|$ for EPR-allowed transitions of group-V donors in silicon ($^{31}$P, $^{75}$As, $^{121}$Sb, $^{123}$Sb and $^{209}$Bi) are plotted as the line thickness in Fig. \ref{fig7}. One can notice that for a given EPR transition, the high-field limit of $\partial \nu / \partial A$ is exactly $m_I$ and, as a consequence, the field sensitivity to the hyperfine interaction is simply written as 
\begin{equation}
\frac{\delta B_z}{\delta A} = \frac{h}{g_e \, \mu_e } m_I.
\end{equation}
Also no polarization of the donor spins is required for SDR spectroscopy; only parallel spin pairs remain in the steady state under illumination. However, at low magnetic field, the donor eigenstates are not pure spin states. Thus, for one transition, the fraction of parallel and antiparallel electron spins of an SDR pair modified by magnetic resonance depends on the magnetic field.\cite{Mortemousque2012}
With such considerations taken into account,  the simulated SDR signal intensity for cw-SDR spectroscopy is plotted by the color scale in Fig. \ref{fig7}.

%%%%%%%%%%%%%%%%%%%%%%%%%%%%%%%%%%%%
%%%%%%%%%%%%%%%%%%%%%%%%%%%%%%%%%%%%
%%%%%%%%%%%%%%%%%%%%%%%%%%%%%%%%%%%%
\section{Summary and conclusions}

In summary, we have performed the cw SDR spectroscopy of \pureSi :Bi and \natSi :Bi at 9 and 7 GHz and observed a significant SDR line narrowing at the HCT. 
The theoretical model proposed in this study for the SDR pair electron distribution reproduces the experimentally obtained line shapes very well. 
%We have modeled the the SDR pair electrons distribution using molecular orbital theory for the different spin configurations. 
By analyzing the line shape at the HCT, we have shown that the main broadening process in \pureSi :Bi is the dipole-dipole interaction between the bismuth donor and the surrounding readout centers.
Our results illustrate fundamental properties of hyperfine clock transitions and serve as a stepping stone for further investigations of coupling between microwave circuits and donors in silicon.
%The paramagnetic centers probed by SDR spectroscopy are subject to the interaction with their readout centers, as well as to intrinsic EPR line broadening processes such as donor- $^{29}$Si hyperfine interaction,\cite{Abe2010} dipolar interaction with similar paramagnetic center\cite{Morley2010,Wolfowicz2012} or even silicon isotope mass composition.

%
%The simulation of $\abs{\partial f/\partial A_c}$ for EPR-allowed transitions is plotted as the line thickness in Fig. \ref{fig7} for the group-V donors in silicon ($^{31}$P, $^{75}$As, $^{121}$Sb, $^{123}$Sb and $^{209}$Bi). As a general feature, $\abs{\delta B_z/\delta A_c}$  for all lines have a non-zero asymptotic value. Phosphorus donor does not have any electric optimal working point for EPR transitions. Other donors have $2I-1$ electric optimal working points. 
%Fig. \ref{fig7} also shows the expected SDR signal intensity\cite{Mortemousque2012} for cw-spectroscopy as a color scale. Red points have a high SDR signal intensity whereas the sample photoconductivity is not changed by magnetic resonance for black ones points. This explain the poor signal to noise ratio for the spectrum of Fig. \ref{fig1}(b) for it has a low expected SDR signal intensity.
%
%\cite{Abe2004}
%\cite{Lepine1972}
%
%

%%%%%%%%%%%%%%%%%%%%%%%%%%%%%%%%%%%%
%%%%%%%%%%%%%%%%%%%%%%%%%%%%%%%%%%%%
%%%%%%%%%%%%%%%%%%%%%%%%%%%%%%%%%%%%
\section*{Acknowledgments}

The authors wish to express their appreciation to  Martin S. Brandt, Felix Hoehne, and David Franke for fruitful discussions. This work has been supported in part by the JSPS Core-to-Core program, in part by MEXT, and in part by FIRST. S.Berger was supported by a JSPS Fellowship for his stay at Keio University. We also acknowledge the Australian Government's NCRIS/EIF programs for access to Heavy Ion accelerator Facilities at the Australian National University.

%%%%%%%%%%%%%%%%%%%%%%%%%%%%%%%%%%%%%%%%%%%%%%%%%%%%%%%%%%%%%%%%%%
% Bibliography
%%%%%%%%%%%%%%%%%%%%%%%%%%%%%%%%%%%%%%%%%%%%%%%%%%%%%%%%%%%%%%%%%%
\bibliographystyle{apsrev4-1}
%\bibliography{biblio}% Produces the bibliography via BibTeX.

%merlin.mbs apsrev4-1.bst 2010-07-25 4.21a (PWD, AO, DPC) hacked
%Control: key (0)
%Control: author (72) initials jnrlst
%Control: editor formatted (1) identically to author
%Control: production of article title (-1) disabled
%Control: page (0) single
%Control: year (1) truncated
%Control: production of eprint (0) enabled
%

%\bibitem [{\citenamefont {Dreher}(2013)}]{DreherPhD}%
%  \BibitemOpen
%  \bibfield  {author} {\bibinfo {author} {\bibfnamefont {L.}~\bibnamefont
%  {Dreher}},\ } Ph.D. thesis, \bibinfo{school}{Technische Universit{\"a}t M{\"u}nchen}, \bibinfo {year} {2013}\BibitemShut {NoStop}%

\end{document}